\documentclass[10pt, technote]{IEEEtran}

\ifCLASSINFOpdf
\usepackage[pdftex]{graphicx}
\else
  \usepackage[dvips]{graphicx}
\fi
\usepackage{epstopdf}

\usepackage[cmex10]{amsmath}
\usepackage{array}

 \usepackage[caption=false,font=footnotesize]{subfig}
\usepackage{fixltx2e}
 \usepackage{color}
     \usepackage{xcolor}

\begin{document}
%
\title{Scanning characteristics of metamirror antennas with sub-wavelength focal distance}
%
%
%

\author{Svetlana~N.~Tcvetkova,
        Viktar~S.~Asadchy,
        and~Sergei~A.~Tretyakov,~\IEEEmembership{Fellow,~IEEE}
\thanks{The authors are with the Department of Radio Science and Engineering, Aalto University, FI-00076 Aalto, Finland (e-mails: svetlana.tcvetkova@aalto.fi; viktar.asadchy@aalto.fi%
).}

}

\maketitle

\begin{abstract}
We investigate beam scanning by lateral feed displacement in novel metasurface based reflector antennas with extremely short focal distances. Electric field distributions of the waves reflected from the antenna are studied numerically and experimentally for defocusing angles up to 24$^\circ$. The results show that despite their sub-wavelength focal distances, the scanning ability of metamirrors is similar to that of short-focus reflectarrays (focal distance about several wavelengths). In addition to offering a possibility to realize extremely small focal distances, metamirror antennas are practically penetrable and invisible for any radiation outside of the operating frequency range.

\end{abstract}

\begin{IEEEkeywords}
Beam-scanning, feed displacement, reflectarrays, metasurfaces.
\end{IEEEkeywords}

%
\IEEEpeerreviewmaketitle

\section{Introduction}
%
%
%
%

\begin{figure}[htp]
{\par\centering
 \subfloat[]{
        \includegraphics[width=0.23\textwidth]{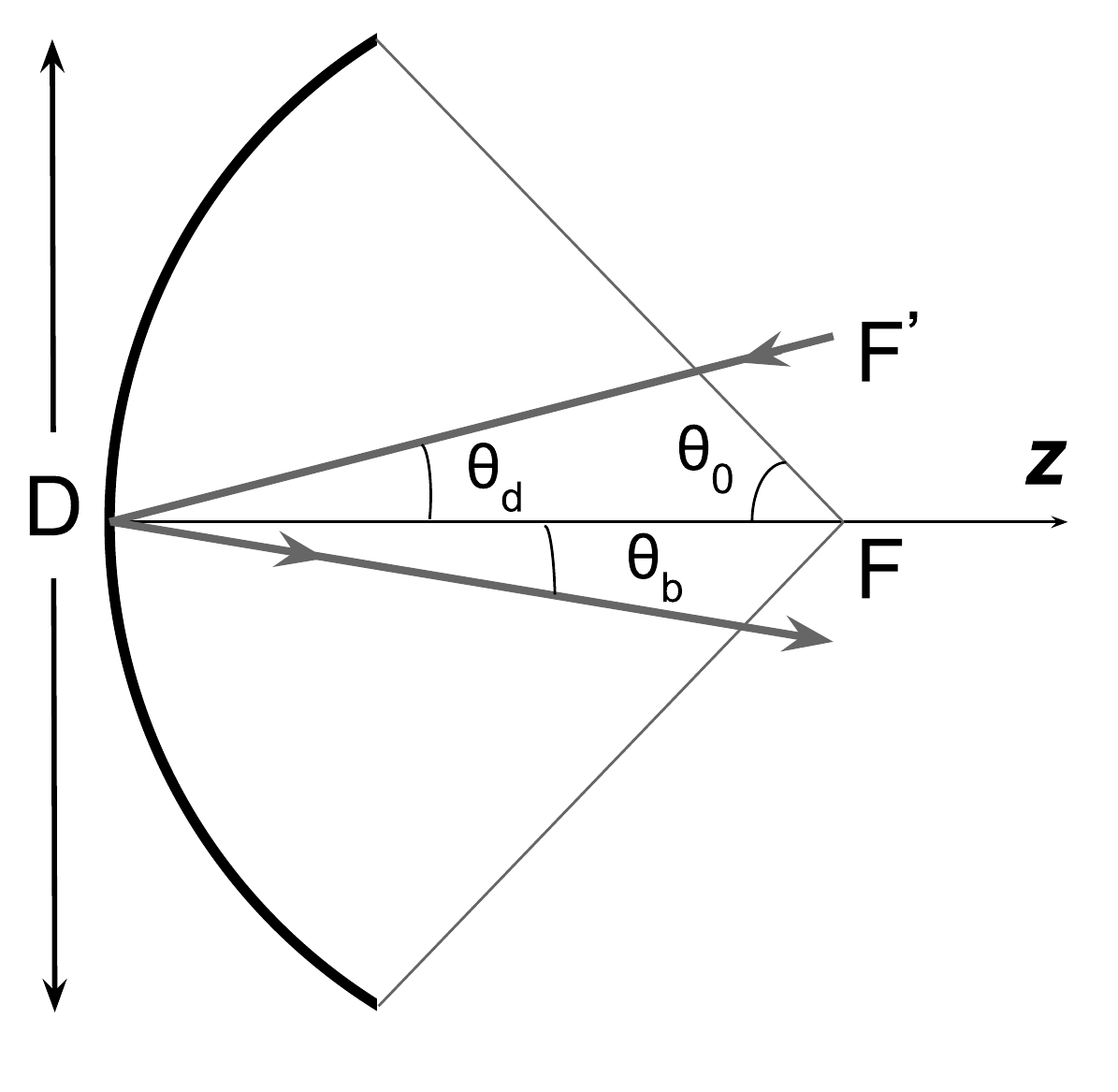}
        \label{fig:ang1}}
 \subfloat[]{
        \raisebox{0.23\height}{\includegraphics[width=0.2\textwidth]{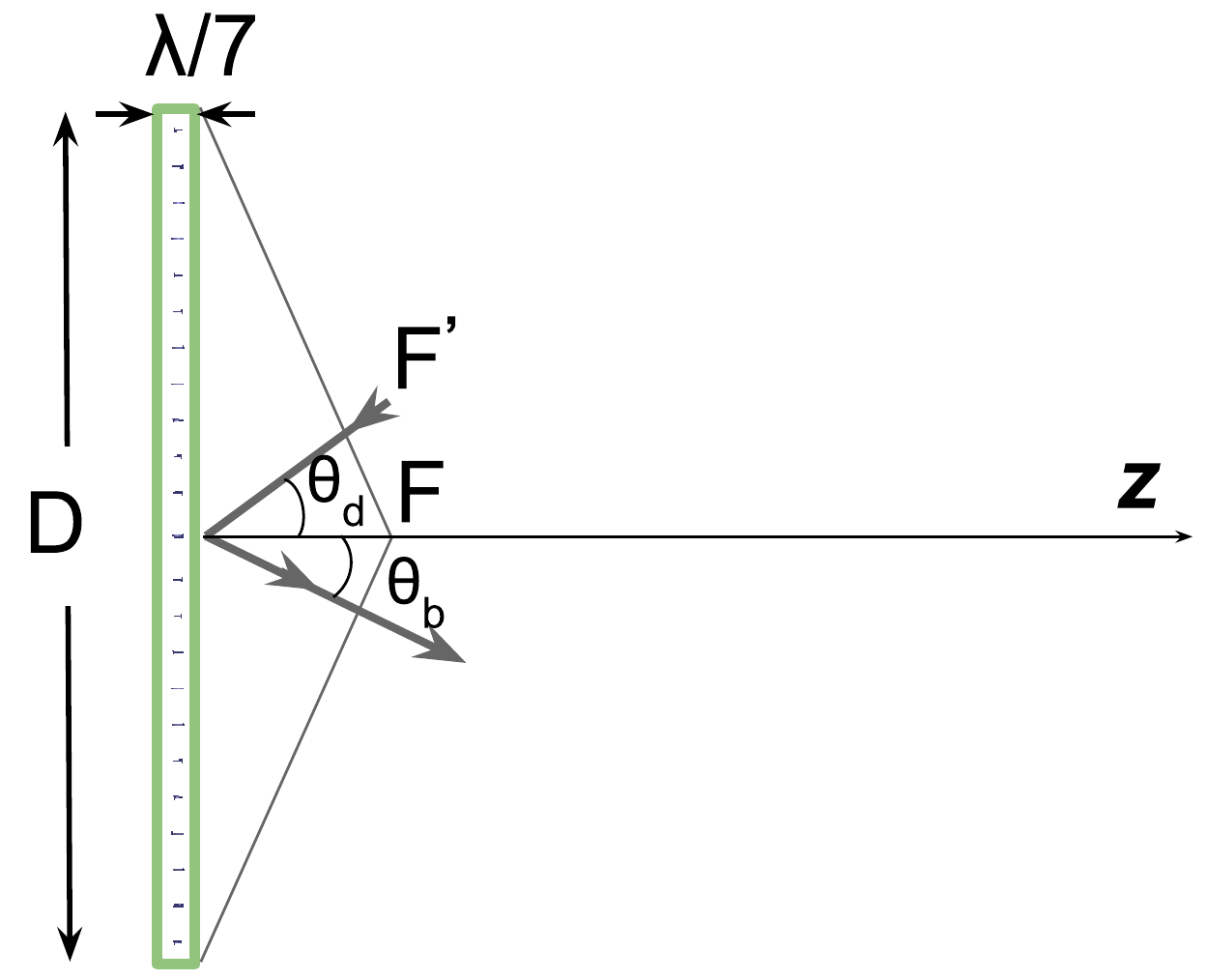}} \label{fig:ang}}
        
  \caption{Lateral feed displacement in reflectors: (a) parabolic antenna, (b) metasurface based reflector}
}\end{figure}

\IEEEPARstart{P}{ossibility} for beam scanning of reflector antennas is an important feature for radar technologies  and other applications. For passive reflectors, beam scanning or producing multiple beams can be accomplished through lateral feed displacement (Fig.~\ref{fig:ang1}).
Geometrical parameters of reflectors are usually defined by the ratio between the focal length and the diameter ($f/D$) or by the subtended angle $\theta_0$ that equals to $\theta_0=\tan^{-1}(0.5 D/f)$ for reflectarrays and $\theta_0=2\tan^{-1}(0.25 D/f)$ for parabolic reflectors \cite{ScanMicrostrip}. 
When the feed is laterally displaced from the focal point of the reflector at a defocusing angle $\theta_d$, the scanned beam is deflected at the angle $\theta_b$.
The beam deviation factor, which can be described as $BDF=\theta_b/\theta_d$ is always less than unity for all known reflector structures.

Conventional reflector antennas, allowing feed-displacement scanning, can be classified into parabolic reflectors and reflectarrays. 
The focal distance of parabolic reflectors generally is much larger than the operational wavelength. Designing reflectors with a short focal distance (to reduce the profile size and weight of the structure) implies parabolic shapes with great depth that compromises the size of the reflector. The mentioned drawbacks of parabolic reflectors can be partially  overcome using planar printed reflectarrays that combine many favourable features of both parabolic reflectors and antenna arrays {\cite{Reflectarray}}. 
Usually, reflectarrays are formed by  arrays of metal patches over a metal ground plane. 
Printed reflectarrays possess such advantages as low profile and light weight, which makes them good candidates for space applications, as well as low price and simplicity of manufacturing. Due to the resonant nature, reflectarrays have limited bandwidth which in most cases does not exceed 15\% \cite{bandwidth,pozar11}. Although the focal distance of reflectarrays can be made considerably smaller than for parabolic reflectors, the smallest realizable values are of the order of several wavelengths.
Many papers \cite{ScanReflectarray,mmwave} describe properties of reflector antennas when the beam-scanning performance is accomplished by lateral displacement of the feed from the focal point. 
Feed displacement can cause substantial degradation of the far-field pattern due to phase aberrations even for reflectors with electrically large focal distances.
Thus, reflector antennas are traditionally designed for limited scan angles because of phase errors which can cause loss in the gain and degradation of the beam \cite{ScanParabola}, although some improved designs are known for the millimeter-waves range {\cite{mmwave}}.

Recently, a new type of reflector antennas was introduced, based on metasurfaces with deeply sub-wavelength thickness \cite{Tailoring,FuncMetamirror} and sub-wavelength focal distances, practically not realizable in other reflectors. These new reflector antennas, called \emph{metamirrors,} are practically transparent and invisible for any radiation with frequencies beyond the desired operational band.
Such a unique functionality can be of paramount importance for many different applications, such as reflector antennas for satellites incorporated with the solar panels (more than 95\% of solar energy is expected to be delivered to the panels through the antenna, as determined by the small geometrical cross section of the metal particles). Another exciting potential application of metamirrors is creating multi-functional and multi-frequency antenna arrays which combine several independent thin layers
operating at different frequencies and performing different functionalities (with the overall thickness still not exceeding the wavelength) \cite{MetamirrorConf}. 
Due to their deeply sub-wavelength constituent elements, metamirrors demonstrate an ability to confine incident energy at an extremely short focal distance, less than the operating wavelength (Fig. \ref{fig:ang}).

However, so far, the scanning characteristics of metamirrors have not been studied. It is known that scanning abilities of conventional reflector antennas (parabolic reflectors and reflectarrays) significantly degrade when the focal distance is reduced \cite{ScanParabola2}.
Thus, it could be expected that metamirrors with sub-wavelength focal distances would experience the problem of extremely high beam degradation when the feed is laterally displaced from the focal point. In reception regime, it could be expected that even a slight deviation of the incidence angle from the target value might lead to severe distortions of the focusing ability. 
In reflection regime, it could appear as a distorted wave instead of a plane wavefront.

The goal of this work is to investigate beam-scanning properties of novel metamirrors with a sub-wavelength focal distance.
We consider two examples of metamirrors emulating spherical and cylindrical lenses in reflection regime. They have the same focal distance but different aperture sizes. We demonstrate, both numerically and experimentally, that despite of the sub-wavelength focal distance, the metasurface based reflectors possess quite acceptable scanning properties suitable for many applications.

%
%

\section{Scanning properties of metamirrors}

\subsection{Ground-free reflector and principles of its operation}

\begin{figure}[!]
\vspace{-0.2cm}
{\par\centering
 \subfloat[]{\includegraphics[width=0.22\textwidth]{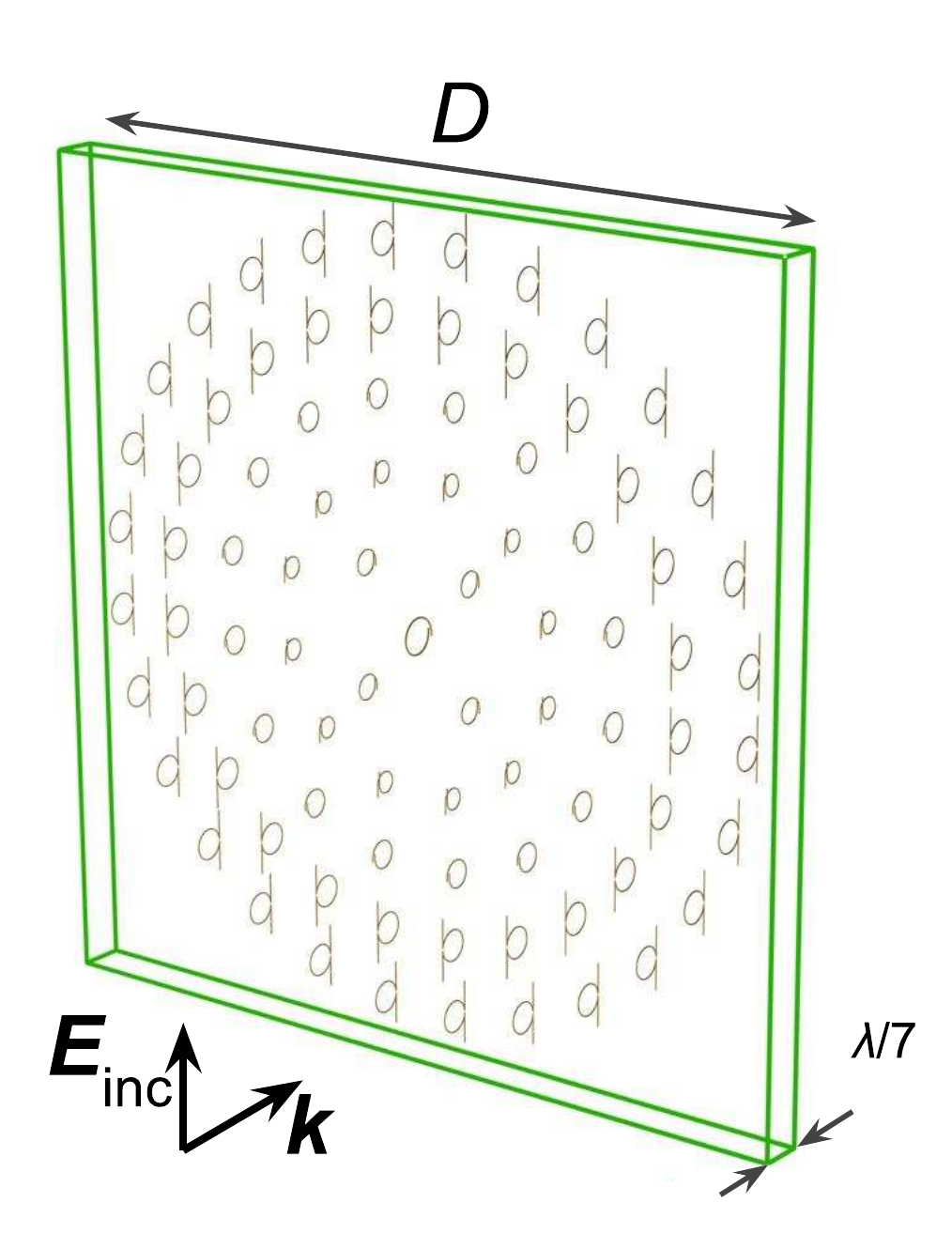}
 \label{fig:planar}}\\
 \vspace{-0.6cm}
 \subfloat[]{\includegraphics[width=0.45\textwidth]{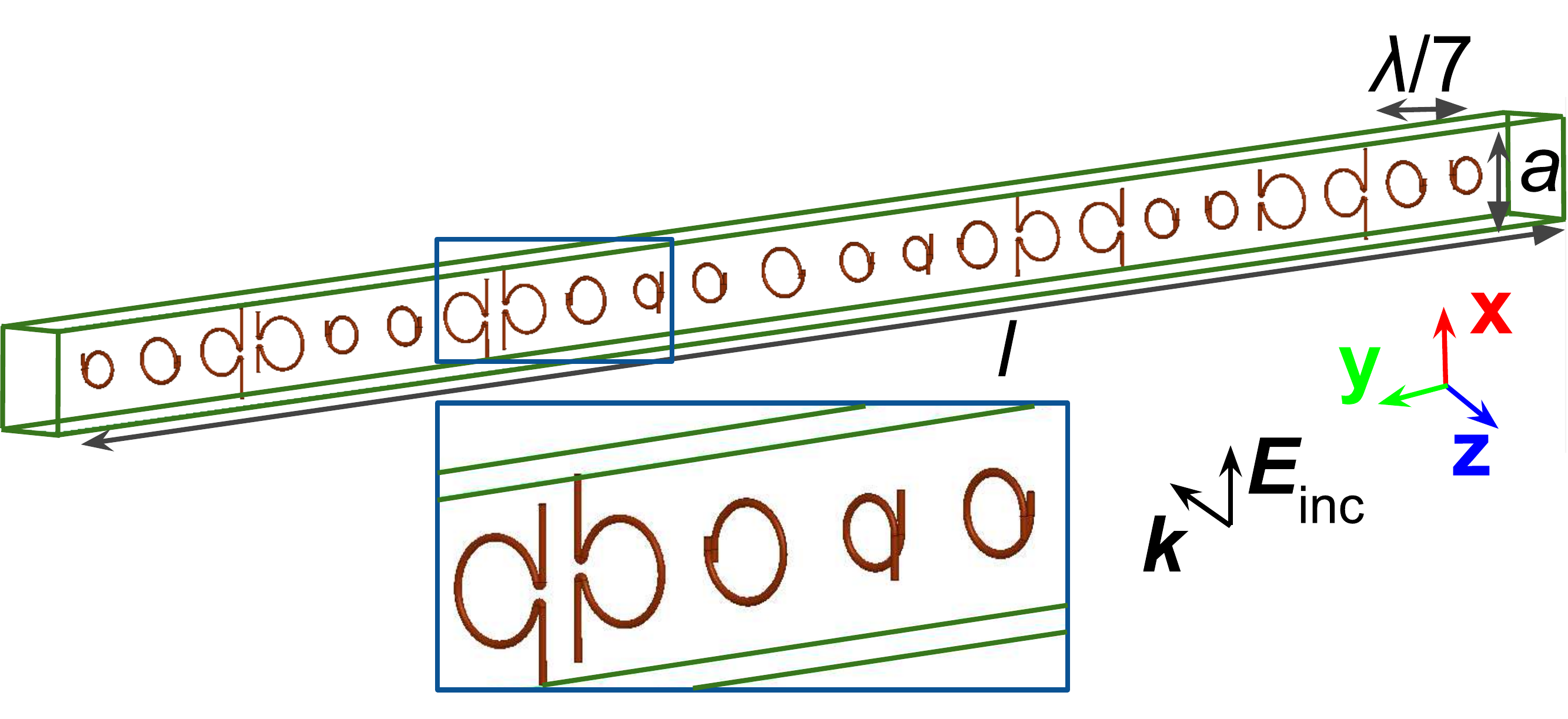}
  \label{fig:linear}}
  \caption{Metamirrors formed by specifically shaped copper elements embedded in a dielectric supporting material layer are marked as a green box. (a) A metamirror emulating a spherical lens in reflection. (b) A single row of the elements of a metamirror imitating a cylindrical lens in reflection.}
 \label{fig:PlanarLinear}}
\end{figure}

The operation mechanism of metamirrors is drastically different from that of reflectarrays and other types of reflector antennas. In contrast to conventional reflectors, a metamirror does not include any ground plane and generally represents a planar array of sub-wavelength sized elements embedded for mechanical support in a low-permittivity dielectric material (see Fig.~\ref{fig:planar}). 
In conventional reflectors, such as reflectarrays, the required phase control is achieved through different sizes or shapes of the patch elements, while the high reflection amplitude is due to a continuous ground plane. Since metamirrors lack a ground plane, their elements perform double functionality. First, they radiate a unit-magnitude wave in the backward direction $\mathbf{\textit E}_{\rm {back}}=e^{j\phi}\,\mathbf{\textit E}_{\rm {inc}}$ with the prescribed non-uniform phase distribution. Second, in the forward direction, they scatter a plane wave (with a uniform phase) which is out-of-phase with the incident wave $\mathbf{\textit E}_{\rm forw}=-\mathbf{\textit E}_{\rm {inc}}$ to ensure their destructive interference and zero transmission. Such asymmetric scattering properties \cite{FuncMetamirror} require both electric and magnetic currents induced in the elements. It was shown in \cite{Tailoring} that the obvious choice, when the elements are realized as simple separated electric and magnetic dipole antennas, is not practical and disadvantageous. It follows from the fact that in this case the electric and magnetic dipole antennas should resonate at different frequencies with different specified amplitudes and be precisely tuned taking into account also their mutual parasitic interactions.

A more effective and practical way to accomplish the required asymmetric scattering from a ground-free reflector is based on the use of small scatterers comprising both induced electric and magnetic currents in a single element. Such magnetoelectric elements comprise electrically polarizable straight wires which are connected to magnetically polarizable wire loops (see Fig.~\ref{fig:linear}) {\cite{OmegaDesign}}. 
Changing the ratio between the length of the straight wires and the diameter of the loop, it is possible to design an arbitrary phase from 0 to $2\pi$ of the backscattered wave from the element (see supplementary material for \cite{FuncMetamirror}).

In metamirrors the phase of backscattering from each element is adjusted individually in order to effectively manage wavefronts of reflection from the structure, maintaining the reflected wave amplitude at the unity value and the necessary phase of the wave scattered in the forward direction.

\subsection{Metamirror emulating a spherical lens in reflection}
We consider a metamirror (Fig.~\ref{fig:planar}) composed of 6 concentric arrays of sub-wavelength elements embedded in a dielectric support with $\epsilon_r=1.03$ and $\tan \delta=0.0001$ \cite{FuncMetamirror}.
Design of such reflector requires certain parabolic phase variation along the surface defined through the equation $\phi(r)=\phi_0+{\omega / c} \,\sqrt{r^2+f^2}, $ where $c$ is a speed of light, $r$ is a distance of inclusion from the center, $f$ is a focal distance, $\phi_0$ is an additional constant phase that can be chosen arbitrarily. It allows to all the fields scattered from each inclusion to interfere constructively in the desired point.
The diameter and the focal distance of the metamirror are $2.8\lambda$ and $0.65\lambda$, respectively. The sub-wavelength focal distance implies a very small focal length to diameter ratio $f/D=0.23$ and a large subtended angle $\theta_0=65^\circ$.  
The phase of reflection for each inclusion from the center to the edges is chosen as follows: $ -80^{\circ}$, $-54^{\circ}$, $0^{\circ}$, $65^{\circ}$, $140^{\circ}$, $-142^{\circ}$.  
The scanning properties of this reflector were studied numerically. The prototype operating at 5~GHz was modelled with a commercial electromagnetic software \cite{hfss}. 

\begin{figure}[h]
{\par\centering
\begin{gather*}{
        \includegraphics[width=0.25\textwidth]{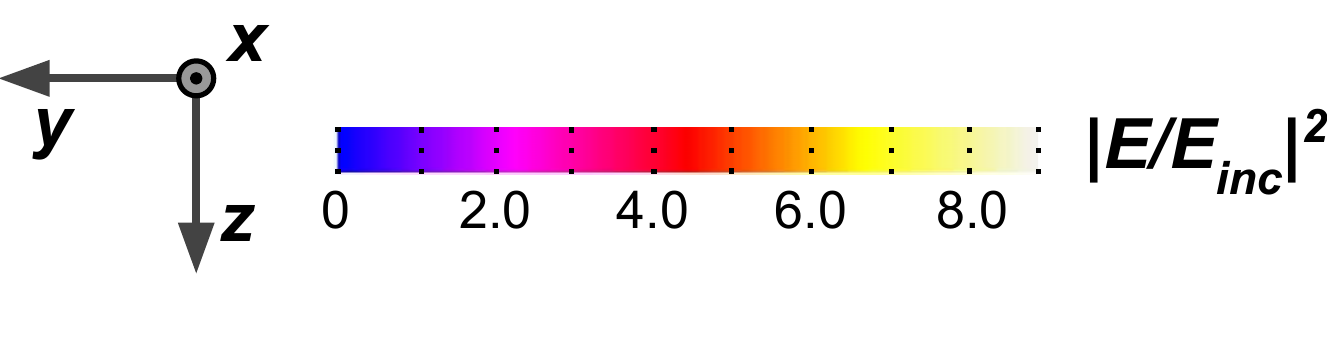}} 
              \end{gather*}\\
              \vspace{-0.5cm}
 \subfloat[]{
        \includegraphics[width=0.10\textwidth, height=4cm]{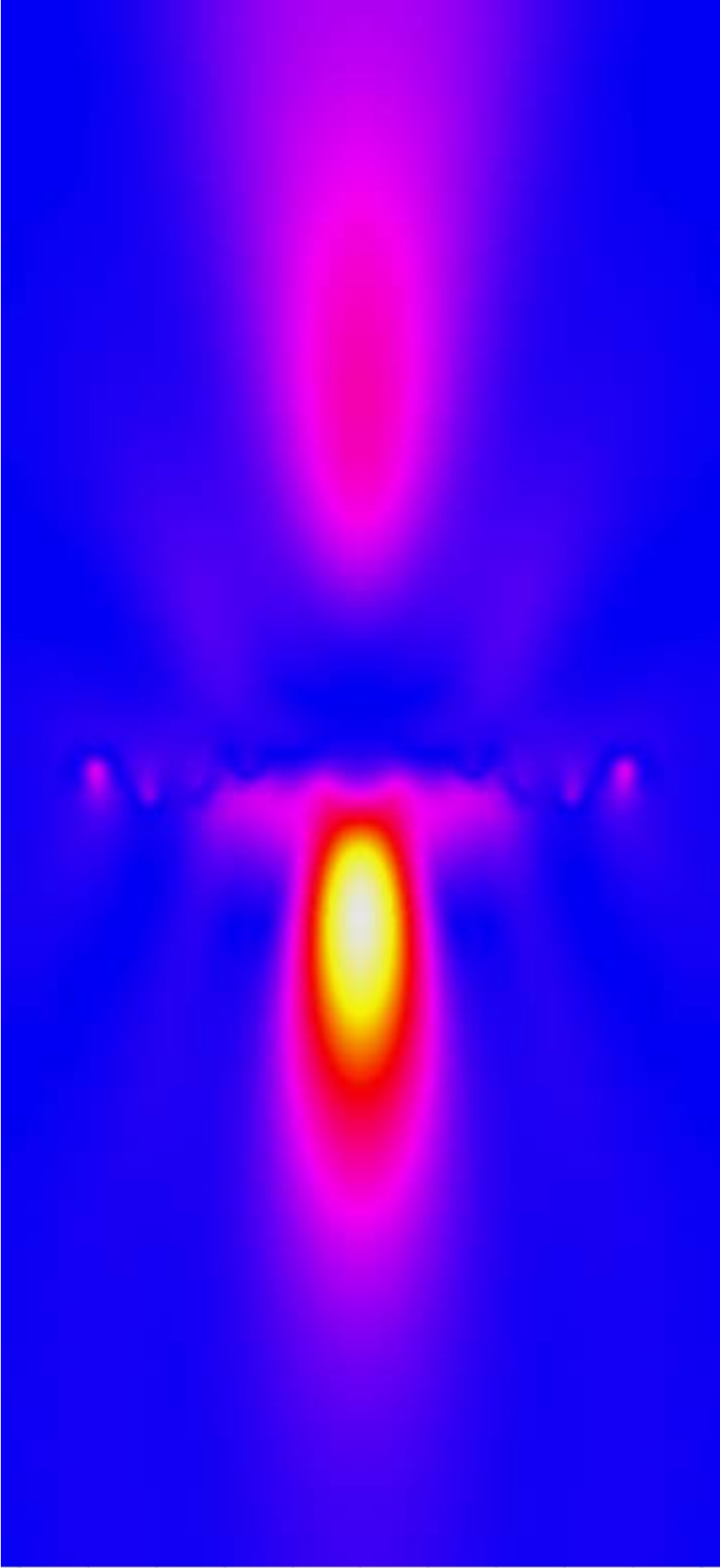}
        \label{fig:0deg}}
 \subfloat[]{
        \includegraphics[width=0.10\textwidth, height=4cm]{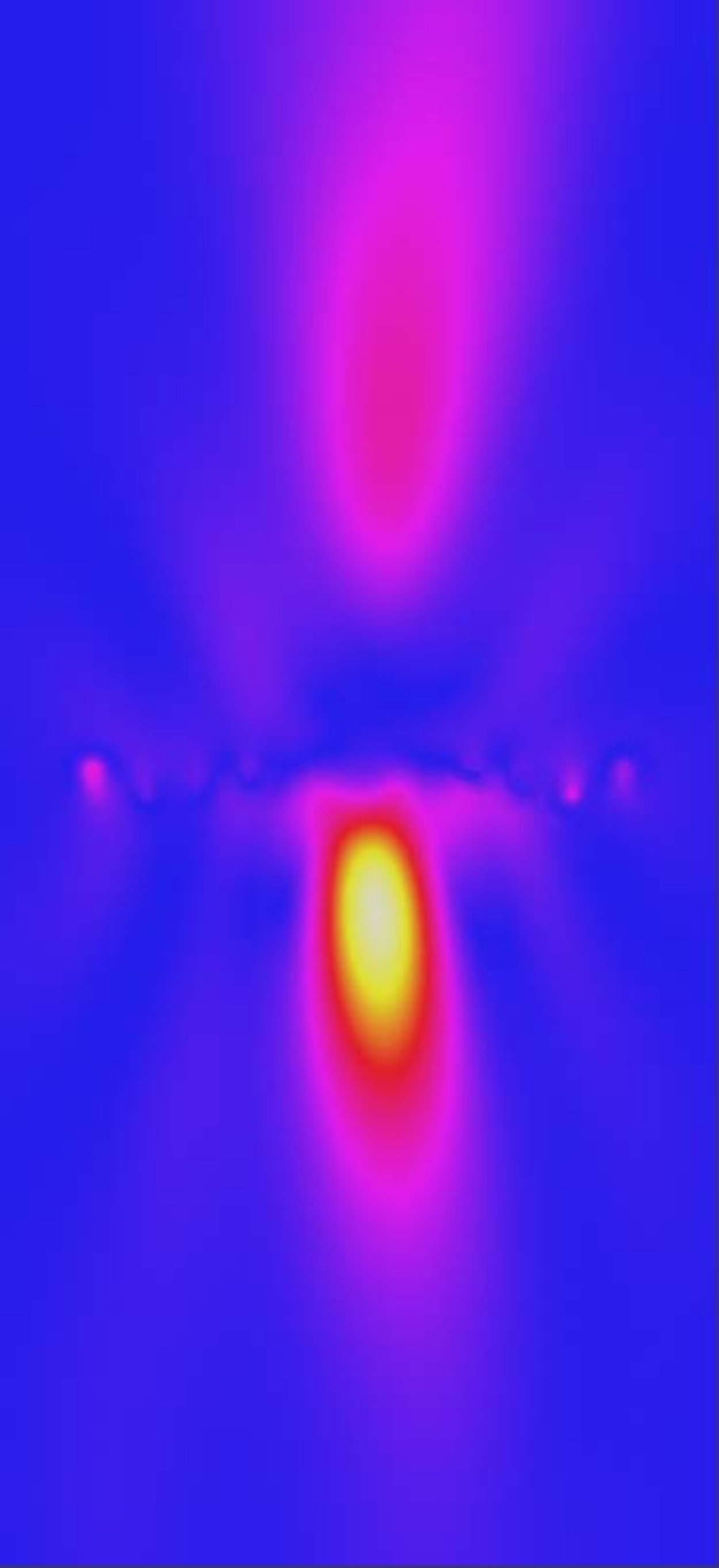}
        \label{fig:5deg}}
 \subfloat[]{
    \label{fig:10deg}
    \includegraphics[width=0.10\textwidth, height=4cm]{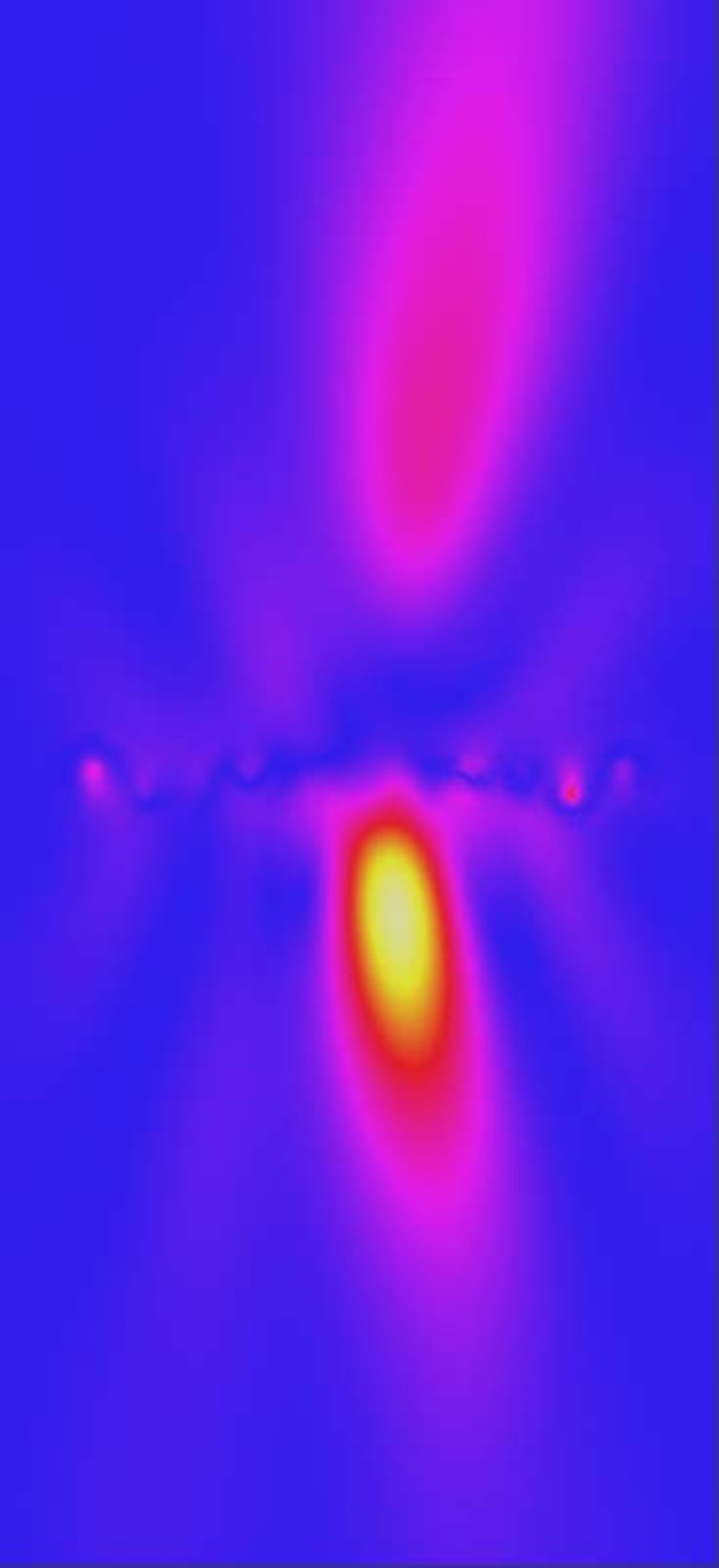}}\hspace{0.03cm}
 \subfloat[]{
    \label{fig:15deg}
    \includegraphics[width=0.10\textwidth, height=4cm]{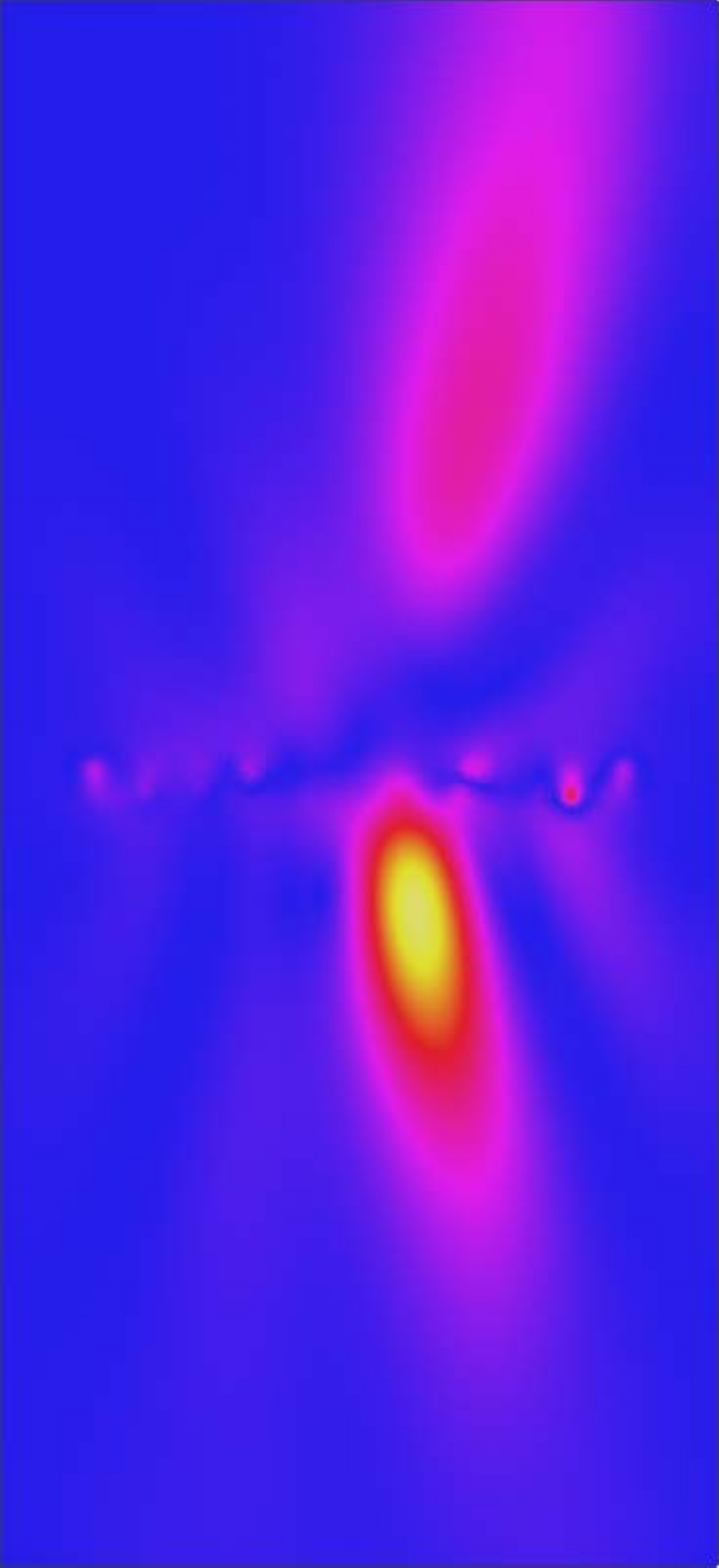}}
 \caption{Simulated results of the power density distribution of the transmitted (the upper $-z$ half-space) and reflected (the lower $+z$ half-space) waves normalized to the incident power density in a case of (from right to left): (a) normal incidence; (b) incident wave at an angle $\theta_b=5^\circ$ from the $+z$-axis (counted towards the $+y$-direction); (c)~incident wave at an angle $\theta_b=10^\circ$; (d) incident wave at an angle $\theta_b=15^\circ$. The metasurface is located in the middle of the plots along the $y$-axis. The incident wave illuminates the metasurface from the bottom.}
 \label{fig:HFSSresults}
}\end{figure}

Based on the reciprocity principle, the metasurface was illuminated by an obliquely incident plane wave and the displacement of the focal spot in reflection was studied. The angle of incidence $\theta_b$ deviated from 0 to $15^\circ$ from the normal ($+z$-direction). 
The power density distributions of the transmitted and the reflected waves normalized to the incident power density for different deviation angles are plotted  in Fig.~{\ref{fig:HFSSresults}}.
The plots show that the metamirror with a sub-wavelength focal distance even at oblique angles efficiently focuses reflected waves with a moderate displacement of the focal spot from the initial position. 
The size of the focal spot in the orthogonal plane is close to that shown in Fig.~\ref{fig:HFSSresults}.
The numerical data are summarized in Table~\ref{tab:planar}. \renewcommand{\arraystretch}{1.3}
\begin{table}[h]
\centering
\begin{tabular}{|c|c|c|c|c|}
\hline
\begin{tabular}[c]{@{}c@{}}Declination \\of the incident \\ wave from \\ the normal\\$\theta_b$, deg\end{tabular} 
& \begin{tabular}[c]{@{}c@{}}Defo -\\cusing\\ angle \\$\theta_d$, deg \end{tabular}  
& \begin{tabular}[c]{@{}c@{}}Power gain\\ in the \\shifted\\ focal spot \end{tabular} 
& \begin{tabular}[c]{@{}c@{}}Power gain \\in the \\ positions\\ of the initial\\ focal spot\end{tabular}  
& \begin{tabular}[c]{@{}c@{}}Beam \\deviation\\factor \\ BDF \end{tabular} 
\\ \hline
0   & 0   & 8.8  & 8.8  & - \\ \hline
5   & 7   & 8.4  & 7.6  & 0.71 \\ \hline
10  & 16  & 8.0  & 5.2  & 0.63\\ \hline
15  & 24  & 7.4  & 2.6  &0.63\\ \hline
\end{tabular}
\caption{Focal spot displacement for obliquely incident illumination}
\label{tab:planar}
\end{table} 
One can notice that the gain in the shifted focal spot becomes slightly smaller when the incidence angle increases, as expected.
Taking into account the extremely short focal distance, the ability of the studied metamirror to collect electromagnetic energy can be considered as remarkable.

\subsection{Metamirror emulating a cylindrical lens in reflection}
In this section we consider another metamirror which focuses the reflected wave in a line parallel to the $x$-axis.
The array is axially symmetric with respect to the $z$-axis. 
Such a structure, operating as a cylindrical lens in reflection, can be easily fabricated and tested experimentally. 
One horizontal row of the metamirror elements is depicted in Fig.~\ref{fig:linear}. 
The metamirror is infinite and periodical along the $x$-axis, while it comprises 23 elements along the $y$-axis with the total length 370~mm. 
Each element is located in a unit cell with the size of $a=15$~mm. 
The thickness of the metamirror is approximately $\lambda/7$ at the operating frequency 5~GHz. 
The focal distance of the structure is $f=0.65\lambda=39$~mm that implies a very small $f/D=0.12$ ratio and a very large subtended angle $\theta_0=77^\circ$.  
The $f/D$ ratio of the present reflector is extremely small compared to the known reflectarrays. 
For instance, in \cite{ScanReflectarray}, the smallest considered ratio is equal to $f/D=0.24$ and corresponds to the subtended angle of $\theta_0=64^\circ$ and, moreover, the author claims that reflectarrays of larger subtended angles are not practical.
 The phase of reflection for each inclusion from the center to the edges is chosen as follows: $-80^{\circ}$, $-54^{\circ}$, $0^{\circ}$, $65^{\circ}$, $140^{\circ}$, $-142^{\circ}$, $-54^{\circ}$, $27^{\circ}$, $113^{\circ}$, $-153^{\circ}$, $-75^{\circ}$, $19^{\circ}$.

The operation of this metamirror was first examined with full-wave simulations. A single row of the elements was placed between the plates of a parallel-plate waveguide with the following dimensions: the sizes along the $y$- and $z$-axes are, respectively, 80~cm and 90~cm, and the height equals to  $a=15$~mm.
Since the elements of the metamirror can be considered as coupled vertical electric and horizontal magnetic dipoles, based on the image principle, a single row of the elements placed in the waveguide emulates a two-dimensional structure infinite in the $x$-direction. In the waveguide at the operating frequency of the metamirror only transverse electromagnetic (TEM) waves can propagate with the fields orthogonal to the direction of propagation. 

\begin{figure}[!]
{\par\centering
 \subfloat[]{
        \includegraphics[width=0.2\textwidth]{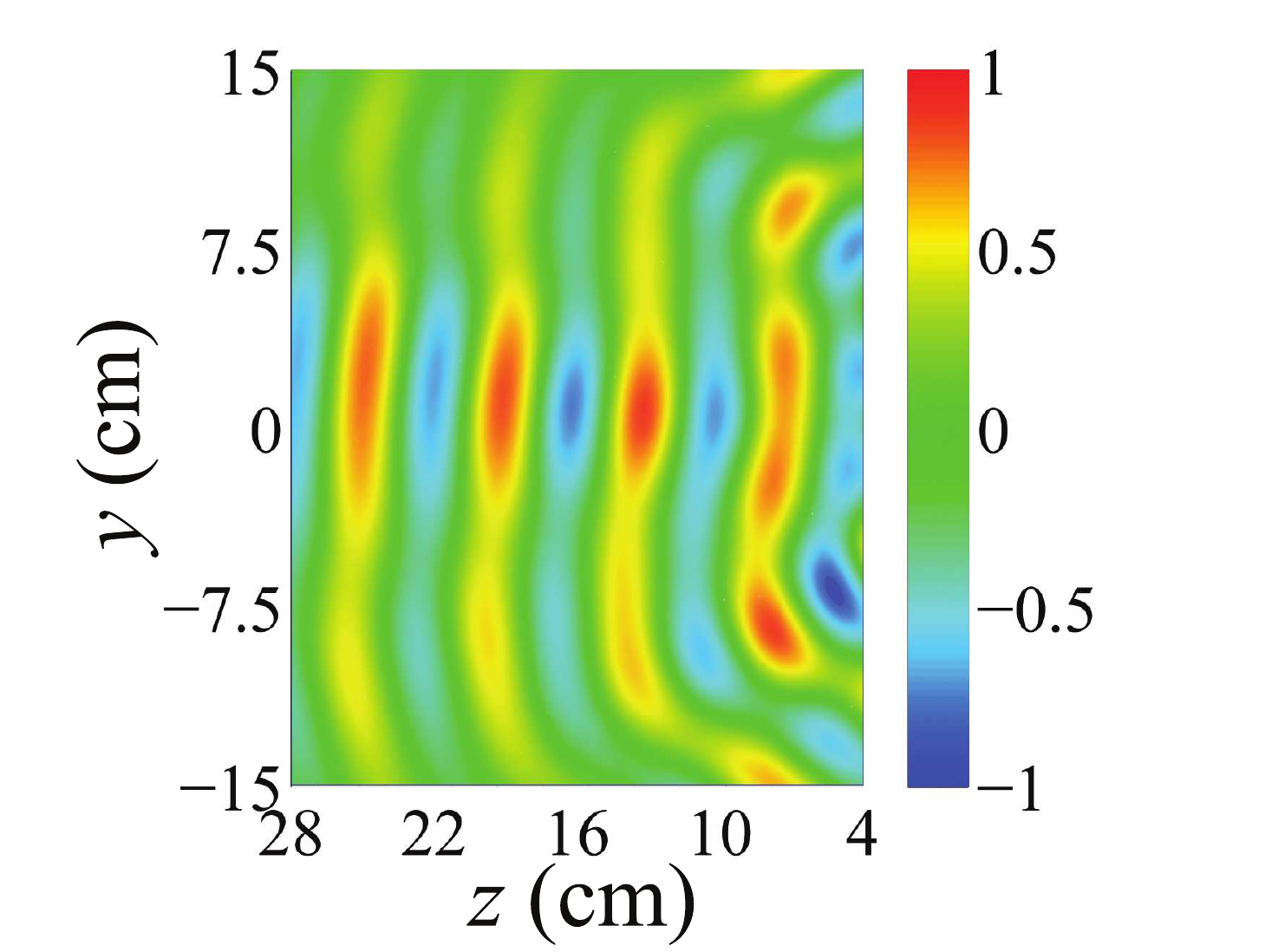}
        \label{fig:5degHFSS}}
 \subfloat[]{
        \includegraphics[width=0.2\textwidth]{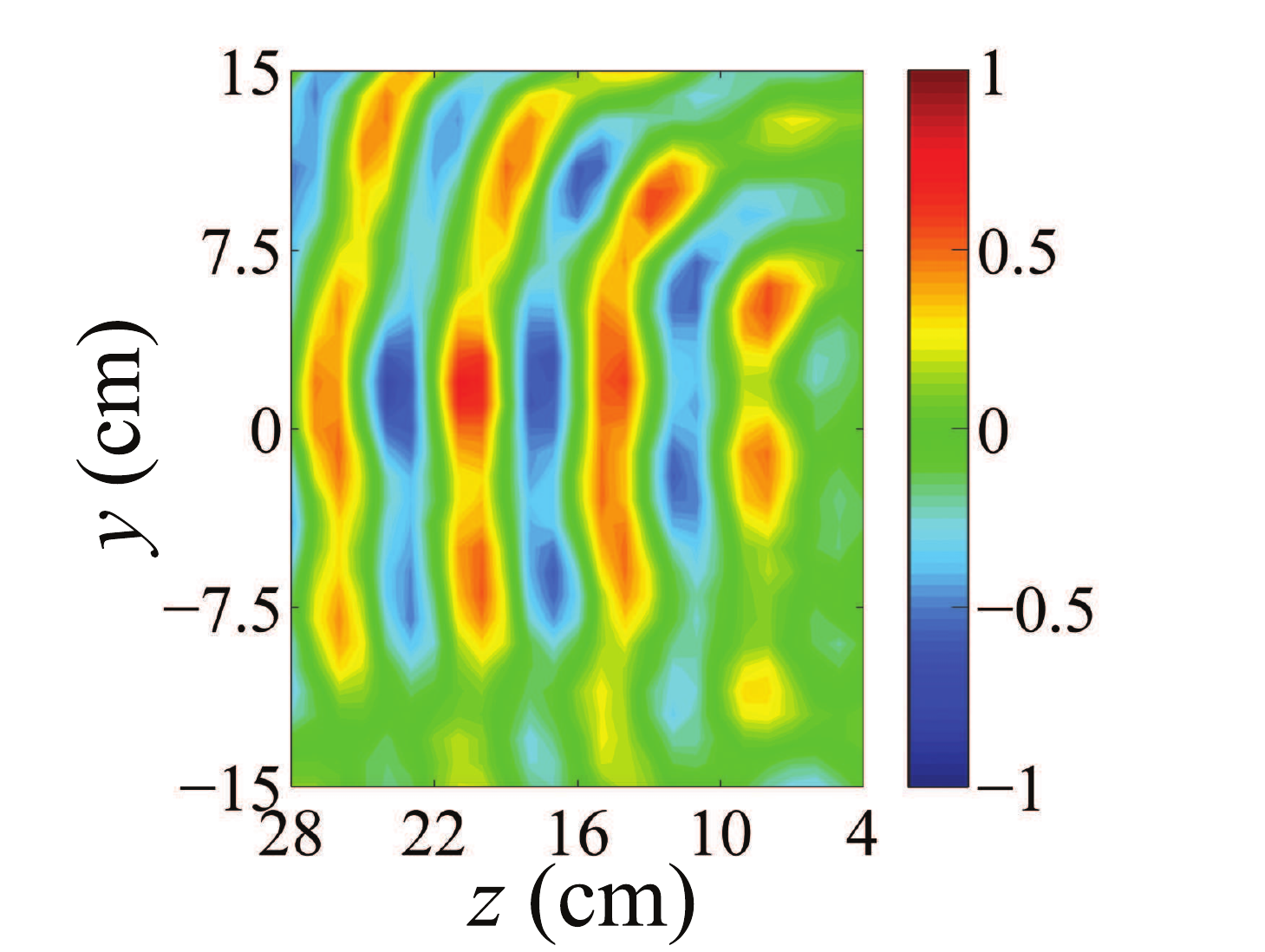}
        \label{fig:5degIT}} \\
                
 \subfloat[]{
    \label{fig:10degHFSS}
    \includegraphics[width=0.2\textwidth]{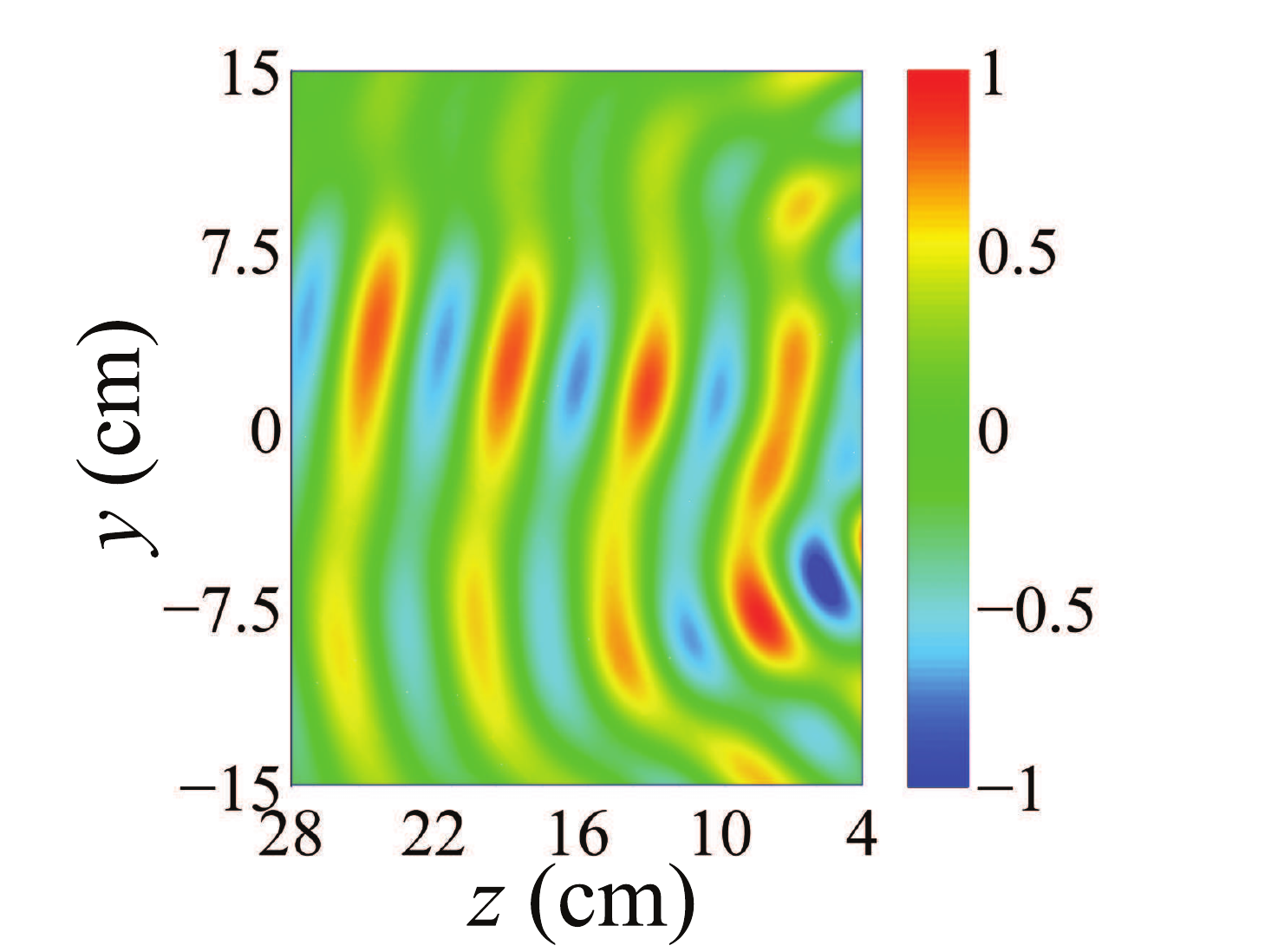}}
 \subfloat[]{
    \label{fig:10degIT}
    \includegraphics[width=0.2\textwidth]{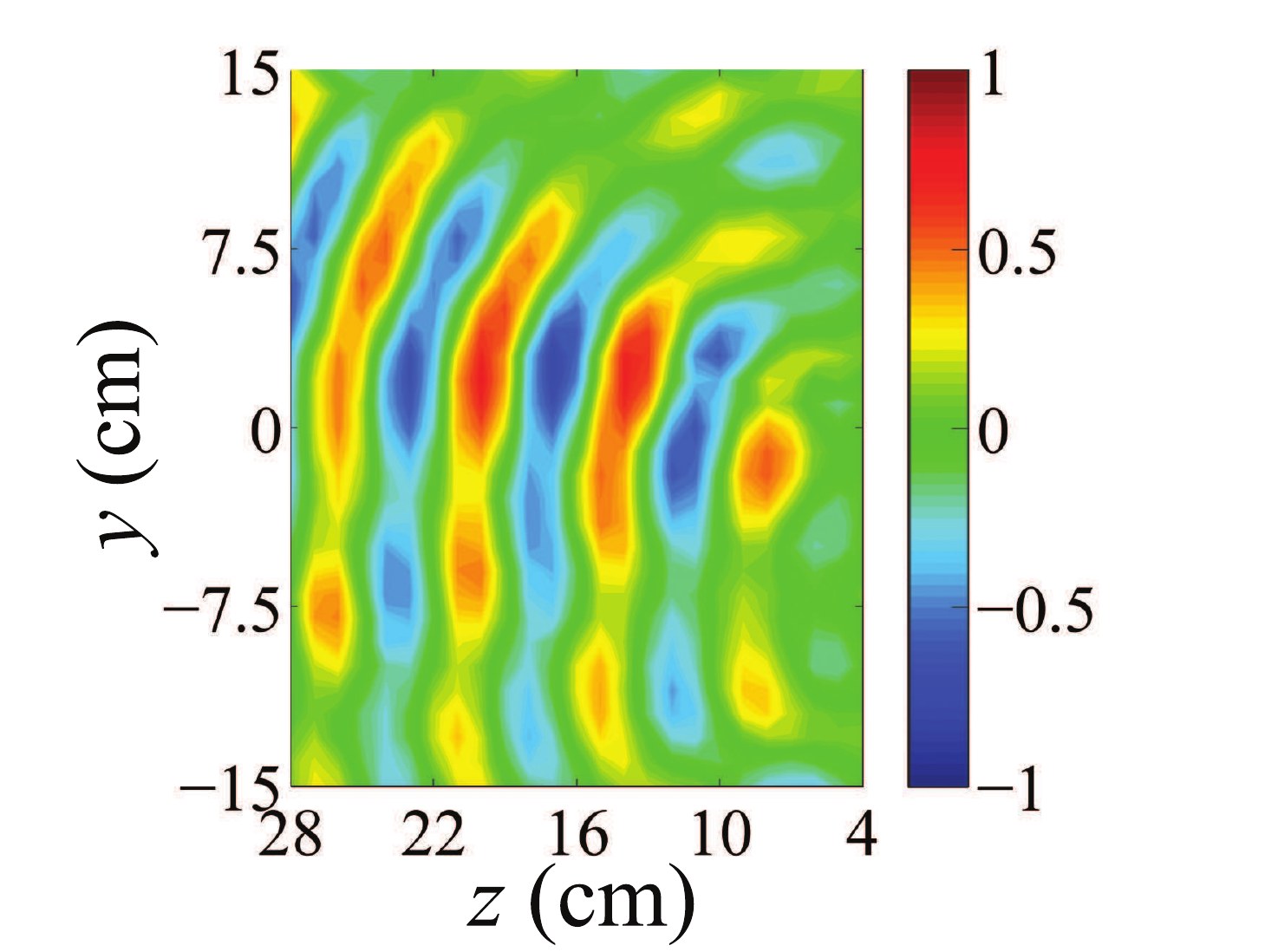}} \\
        
    \subfloat[]{
        \includegraphics[width=0.2\textwidth]{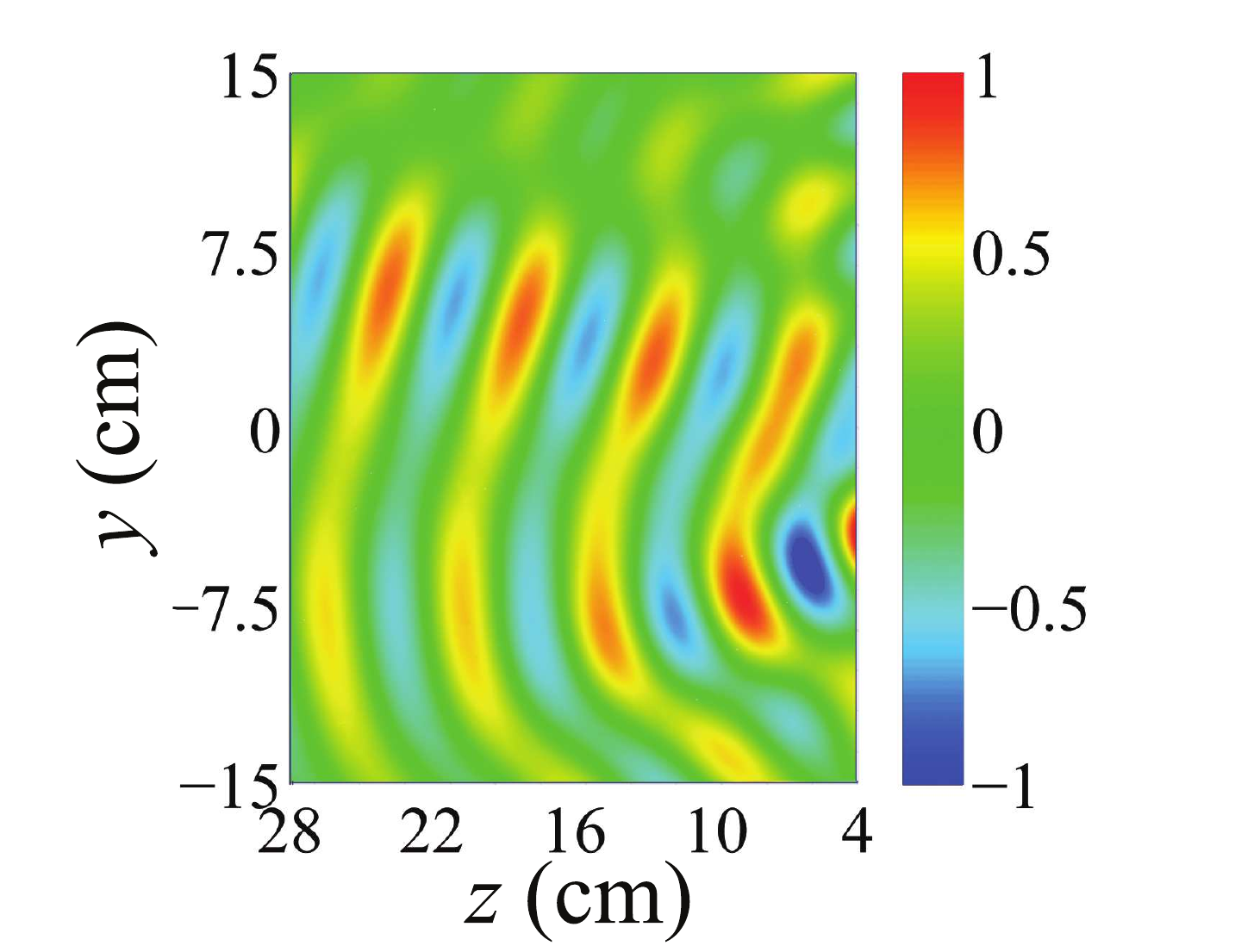}
        \label{fig:15degHFSS}}
     \subfloat[]{
        \includegraphics[width=0.2\textwidth]{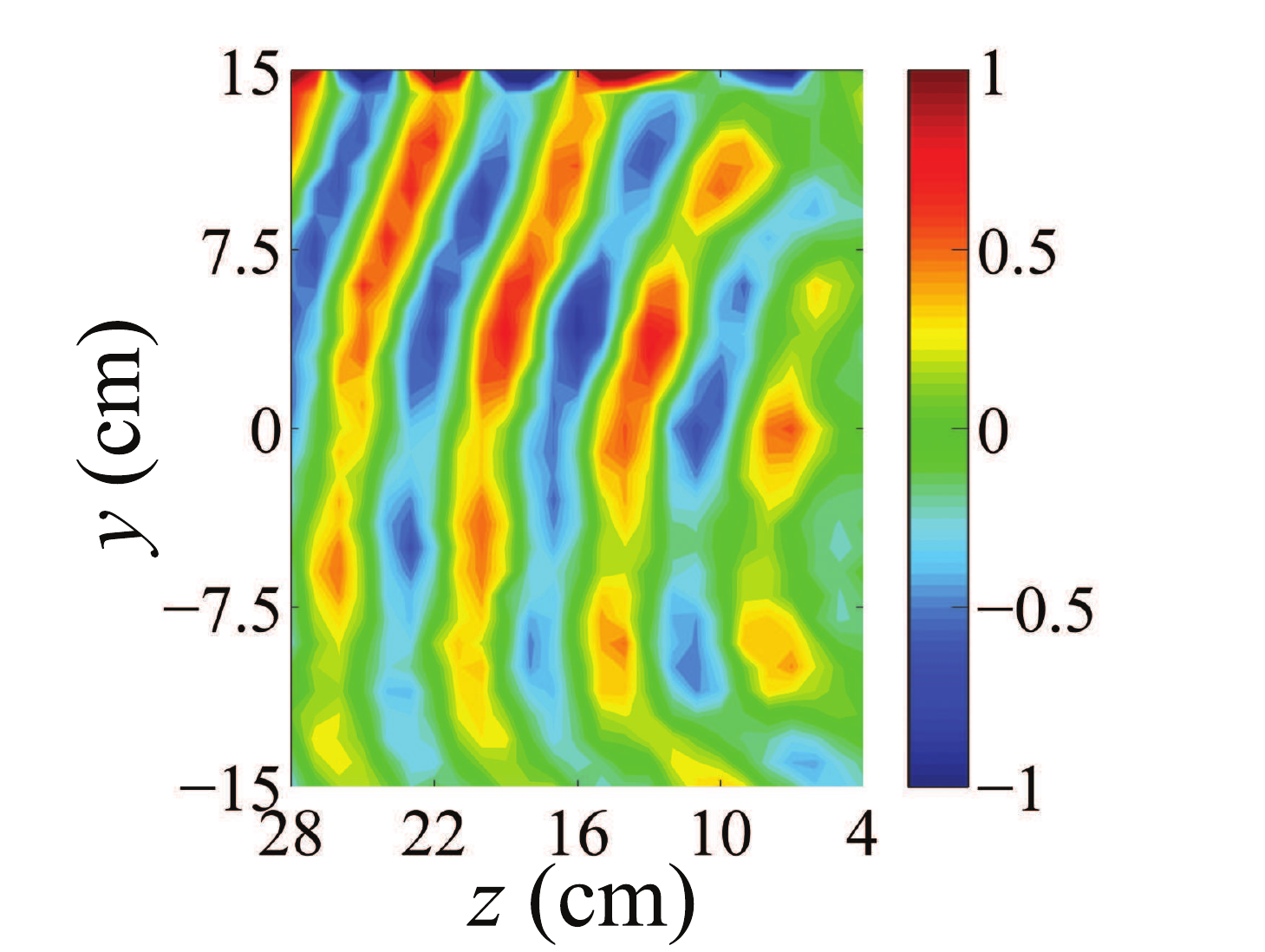}
        \label{fig:15degIT}}
  \caption{The electric field distributions of the wave reflected from the metamirror. The feed and the metamirror are positioned at $z=3.9$~cm and $z=0$~cm, respectively.
Plots (a), (c) and (e) depict the simulated field distributions for the cases when the feed was displaced by  $\theta_d=7^{\circ}$, $\theta_d=16^{\circ}$ and $\theta_d=24^{\circ}$, respectively. Plots (b), (d) and (f) show the corresponding measured field distributions.}
}\end{figure}

The feed in the simulation was a vertical coaxial source that generated a cylindrical wave with the electric field along the $x$-axis. The feed was located at the focal distance of the metamirror on the $z$-axis. 
On the edges of the waveguide the boundary conditions of a perfectly matched layer were applied to avoid parasitic reflections.
The source was displaced at 5~mm, 10~mm and 17.5~mm from the focal point along the $y$-axis, which corresponds to the defocusing angles $\theta_d=7^{\circ}$, $\theta_d=16^{\circ}$ and $\theta_d=24^{\circ}$. 
The simulated results for these three cases are shown in Figs.~{\ref{fig:5degHFSS}, \ref{fig:10degHFSS} and \ref{fig:15degHFSS}}. 
\renewcommand{\arraystretch}{1.3}
	\begin{table}[htp]
		\centering
		\begin{tabular}{|c|c|c|c|c|}
			\hline
			\begin{tabular}[c]{@{}c@{}}Defo-\\cusing\\  angle \\  $\theta_d$, $\deg$ \end{tabular} 
			& \begin{tabular}[c]{@{}c@{}}Lateral  \\ displacement \\ from the\\focal spot,\\ mm \end{tabular} 
			& \begin{tabular}[c]{@{}c@{}}  Scanned \\ beam angle \\ (simulation/\\experiment)\\ $\theta_b$, $\deg$ \end{tabular}  
			& \begin{tabular}[c]{@{}c@{}} BDF \\(simulation/\\experiment) \end{tabular} 
 			& \begin{tabular}[c]{@{}c@{}} Main beam \\ level \\{(simulation)},\\ dB   \end{tabular}  
				\\ \hline
7   & 5 & 3.0 / 4.4 & 0.43 / 0.63 & 16.1   \\ \hline
16   & 11 & 6.0 / 7.0 &  0.38 / 0.44 & 14.7  \\ \hline
24   & 17 & 9.5 / 11.0 & 0.40 / 0.46  & 12.2  \\ \hline
			\end{tabular} \\
		\caption{The beam steering due to the lateral feed-displacement}
		\label{tab:1}
	\end{table} 
The numerical data obtained in simulations are summarized in Table~\ref{tab:1}. Comparing the results in Table~\ref{tab:planar} and Table~\ref{tab:1}, one can notice that the beam deviation factor is smaller for the case of the metamirror with cylindrical symmetry. This can be explained by the fact that the cylindrical metamirror has a greater length that results in increasing  phase errors  and decreasing the main beam steering angular range (stronger displacement of the focal spot for the oblique incidence).  

\begin{figure}[h]
{\par\centering
 \subfloat[]{
      \includegraphics[width=0.3\textwidth]{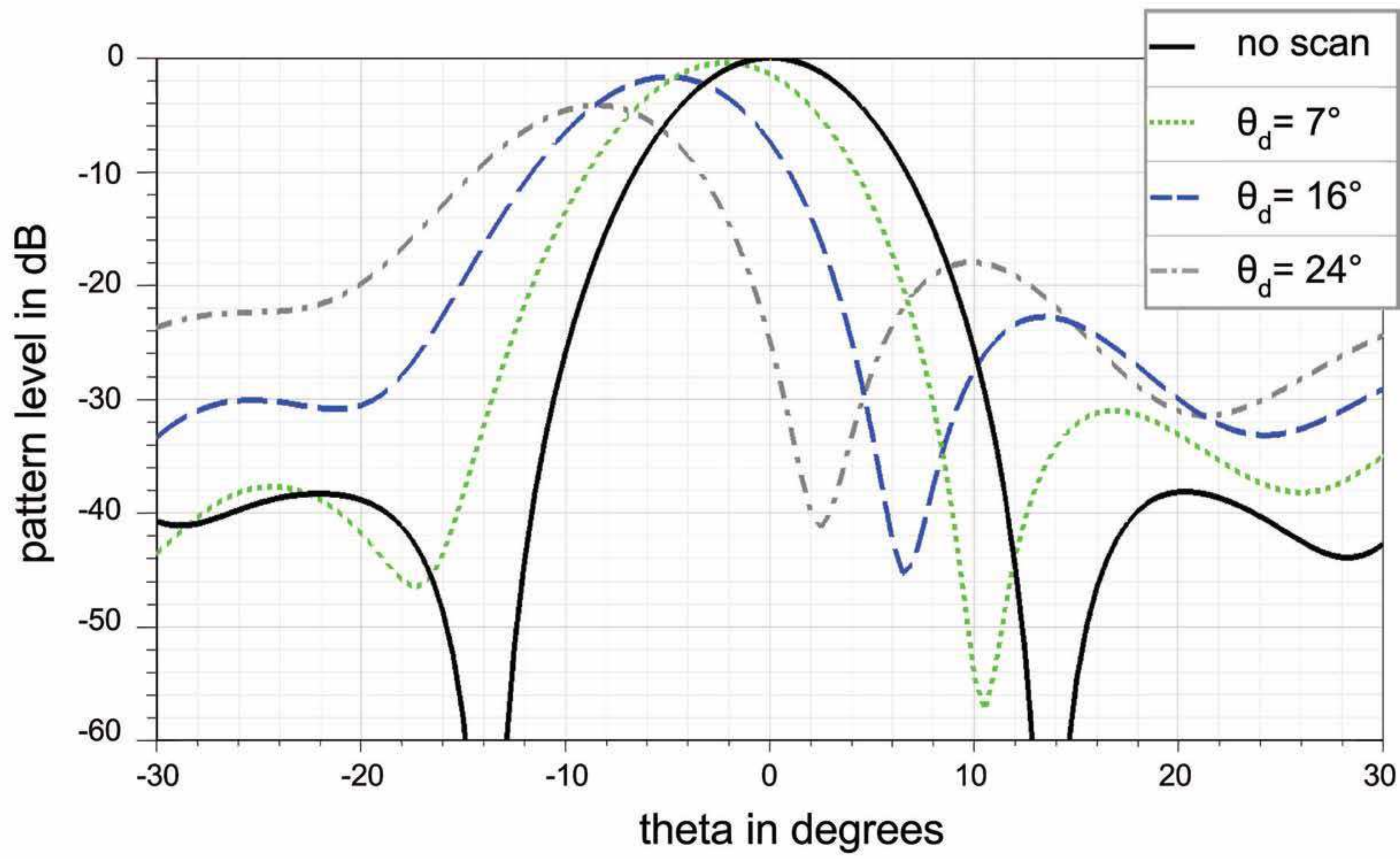}
        \label{fig:pattern1}}\\
\subfloat[]{
 \includegraphics[width=0.4\textwidth]{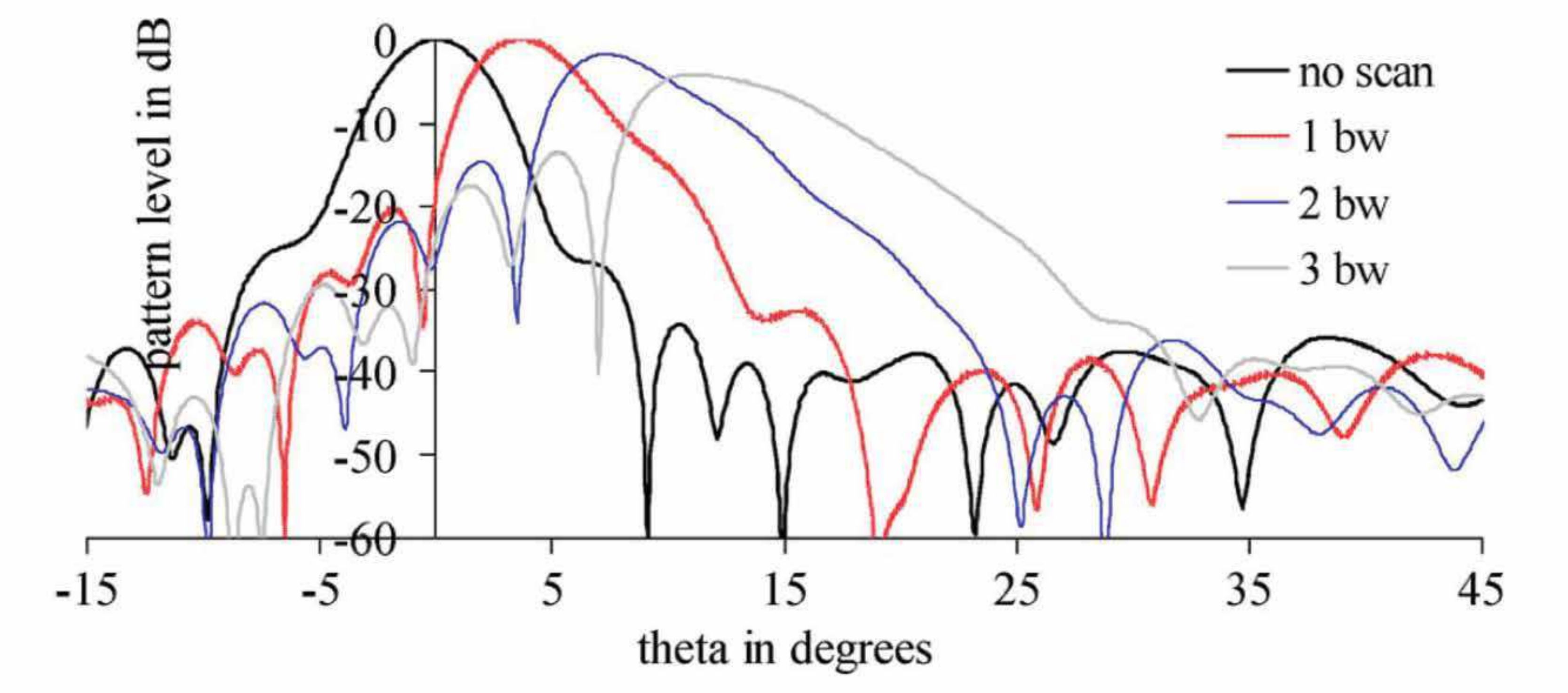}
  \label{fig:pattern2}}
   \caption{ (a) Simulated scanned beam pattern for the metamirror with subtended angle $\theta_0=77^{\circ}$ ($f/D=0.12$)(b) Scanned beam patterns of the broadside microstrip reflectarray when the feed was displaced at a distance equal to 1, 2 or 3 beamwidths of the reflector's main beam ($\theta_0=64^{\circ}$, $f/D=0.24$).}
 \label{fig:reflectarray}
}\end{figure}

The reflection coefficient of the explored metasurface on the operating frequency is equal to $70\%$, while the transmission and absorption reach $20\%$ and $10\%$, respectively. 
Moreover, such structure is penetrable outside of the quite narrow operating frequency range (around $5\%$). For instance, the transmission coefficients for 3 and 5 GHz are $95\%$ and $97\%$, respectively.
 Fig.~\ref{fig:pattern1} shows the radiation patterns of the metamirror in the $yz$-plane.
We can conclude that despite of the sub-wavelength focal length the scanning characteristics of the studied metamirror are at the same level as those of comparable reflectarrays. 
For instance, a microstrip antenna described in paper \cite{ScanReflectarray} has the diameter about $22\lambda$ and operates at 28~GHz. 
The lowest ratio between the focal distance and aperture diameter  discussed by the author is approximately $f/D=0.24$ and it corresponds to the subtended angle equal to $\theta_0=64^{\circ}$. 
Computed radiation patterns of the reflectarray for the cases when the feed was displaced at a distance equal to 1, 2 and 3 beamwidths are depicted in Fig.~\ref{fig:pattern2}. 
Comparison with the results of  \cite{ScanReflectarray}  shows that the scanning performance of the metamirror is on par with the reflectarray despite the fact that its focal distance is several times smaller and the subtended angle larger.

By comparing the two studied metamirrors with different diameters, one can conclude that the number of phase overlaps increases with subtended angle value increase (or reduction of $f/D$ ratio), because the path difference between different points of the reflectarray and the feed can consist of a greater number of wavelengths.
Moreover, the beam deviation factor dependence on the length of the structure was noticed within the same defocusing angles.
The longer array, the smaller scanned beam angles $\theta_b$  can be achieved.

\subsection{Experimental verification}

\begin{figure}[h]
\centering
      \includegraphics[width=0.4\textwidth]{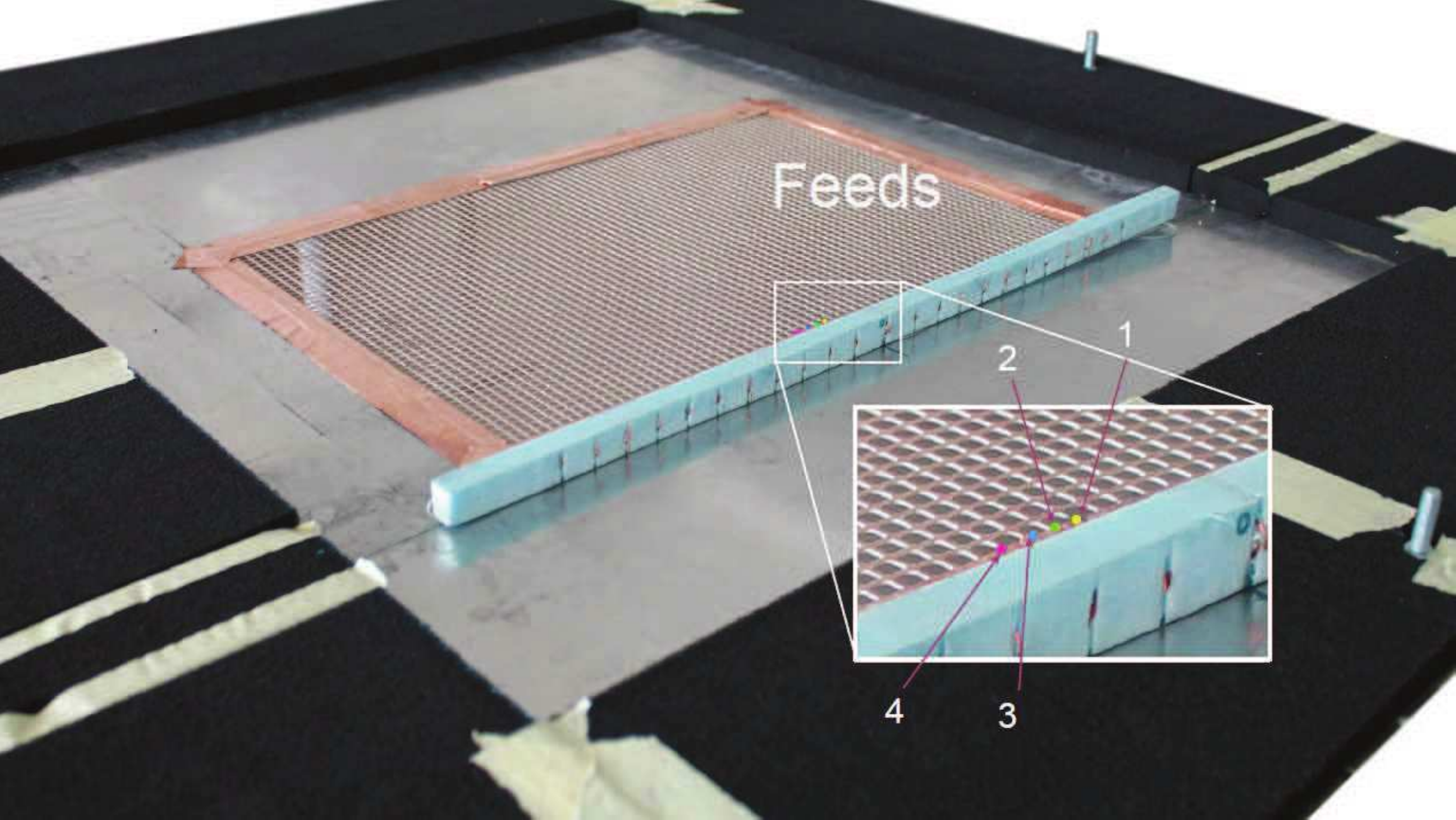}
   \caption{Experimental setup. The bottom part of the waveguide with a copper mesh and the metamirror. Feed positions marked by numbers: 1 -- the initial focal point, 2 -- feed displaced at 5~mm ($\theta_d=7^{\circ}$), 3 -- feed displaced at 11~mm ($\theta_d=16^{\circ}$), 4 -- feed displaced at 17.5~mm ($\theta_d=24^{\circ}$).}
 \label{fig:exp1}
\end{figure}

To verify the scanning properties of the metamirror considered in the previous section, measurements in a parallel-plate waveguide were conducted. 
A vertical coaxial feed was used as a source of incident cylindrical waves with the $x$-oriented electric field. The feed position was changed laterally from the focal point of the metamirror along the $y$-axis (see Fig.~\ref{fig:exp1}).

\begin{figure}[h]
\centering
      \includegraphics[width=0.3\textwidth]{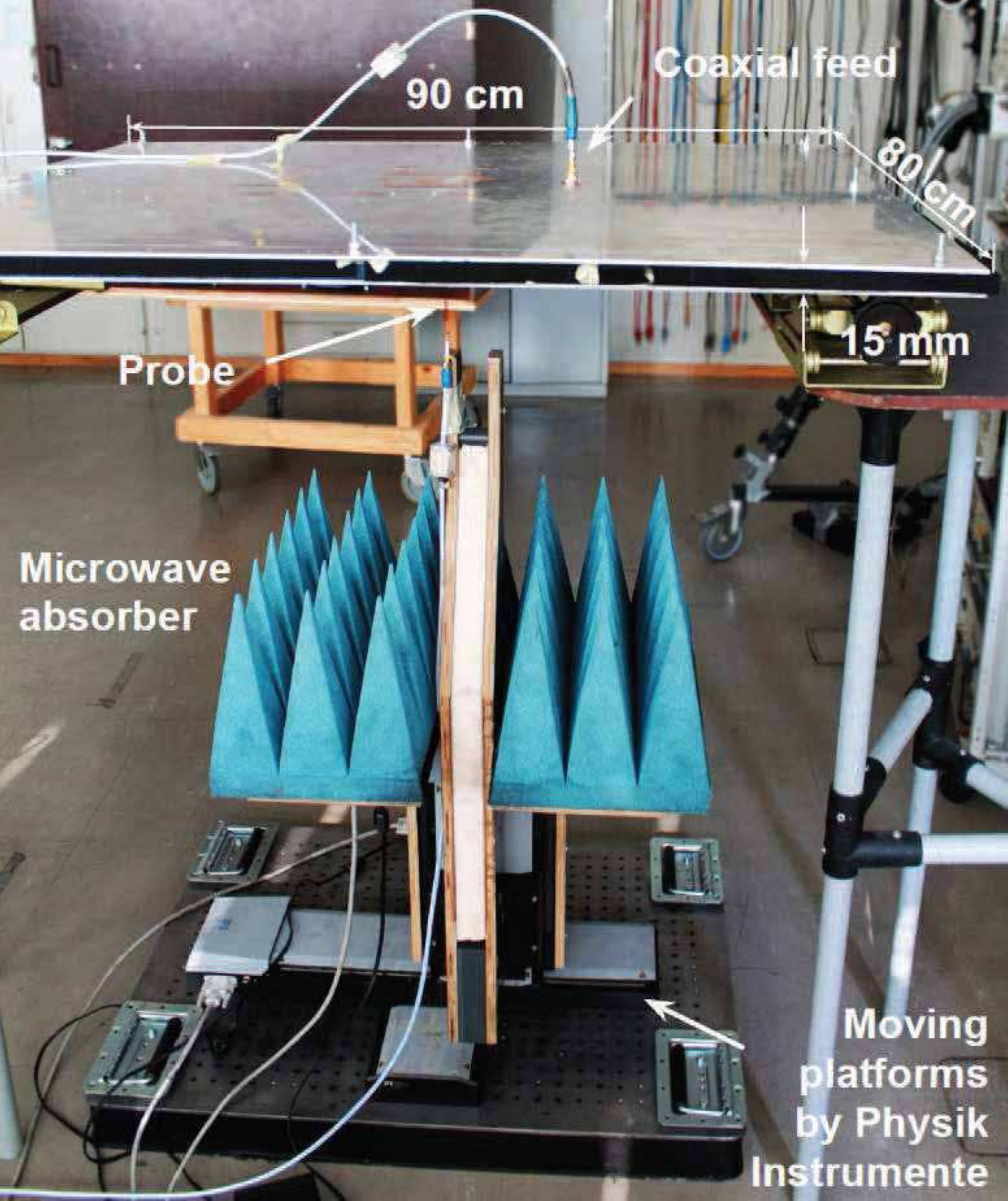}
   \caption{Experimental construction: the moving platforms (Physik Instrumente), coaxial probe antenna, coaxial feed antenna and parallel-plate waveguide.}
 \label{fig:exp_full}
\end{figure}

In the bottom plate of the waveguide a copper mesh ($25\times35$~cm)  was embedded. 
Using a movable coaxial probe positioned under the mesh, it is possible to measure the spatial distribution of the $x$-component of the electric field inside the waveguide (see Fig.~\ref{fig:exp_full}). The mesh period (5~mm) is much smaller than the wavelength, therefore the field distribution inside the waveguide is not significantly disturbed by the mesh. 
A vertical coaxial probe located 5~mm below the mesh gauged the near fields penetrated through the mesh. 
To reduce parasitic reflections from the edges of the waveguide, it was necessary to place microwave absorbing material blocks of 10~cm width at the edges of the waveguide. The measuring device in the experiment is a vector network analyzer Agilent Technologies E8363A. 
Port~1 and port~2 of the analyzer were connected to the stationary coaxial feed and to the movable coaxial probe, respectively. 
The transmission coefficient $S_{21}$ from port~1 to port~2 was measured by the scanning system with the step of 10~mm. 
The spatial distribution of the $x$-component of the electric field can be  represented by the distribution of the transmission coefficient. 

Two sets of measurements are needed to determine the distribution of the reflected fields from the metamirror. 
First, the empty waveguide was analysed to obtain the field distribution of the incident wave. In the second set of measurements with the metamirror placed inside the waveguide, the total fields were measured. 
The reflected fields distribution was found by subtracting the incident fields from the total ones.

The measured field distribution of the reflected wave from the metamirror for all  three cases of feed displacement are presented in  Figs.~\ref{fig:5degIT}, \ref{fig:10degIT} and \ref{fig:15degIT}.
As it is seen from Table~\ref{tab:1}, the scanned angles of the main beam in the experiment are well correlated with the corresponding values achieved through full-wave simulations.

\section{Conclusion}

In this paper we have examined novel metasurface-based reflectors for their scanning ability at feed displacements  and the focusing ability for incident waves deviating from the main-beam direction.  
Although it could have been expected that a reflector with a sub-wavelength focal length might be impractical due to its potentially strong sensitivity to the feed position, both the simulated and measured results of the considered extremely short-focus metamirrors reveal their modest, but acceptable for many applications  scanning properties. 
In fact, the scanning abilities are on par with state-of-the-art short-focus reflectarray antennas, despite the fact that the focal distance of metamirrors is significantly smaller, even smaller than the wavelength. 
Being transparent outside the operational frequency range, single-layered or even cascaded metamirrors can be exploited in various applications, in particular for satellites, for radioastronomy and together with solar energy harvesting systems.

\section*{Acknowledgment}

This work was supported by Academy of Finland (project 287894). The authors would like to thank Dr. E.L. Svechnikov for his useful comments on preliminary results of this work.

\ifCLASSOPTIONcaptionsoff
  \newpage
\fi

\end{document}